\documentclass[pre,twocolumn,aps,floatfix]{revtex4-1}
\usepackage{amsmath}
\usepackage{amsfonts}
\usepackage{amssymb}
\usepackage{graphicx}
\usepackage{fontenc}
\usepackage{psfrag,layout}
\usepackage{color}

\begin{document}

\title{Crystal growth kinetics in Lennard-Jones and Weeks-Chandler-Andersen 
systems along the solid-liquid coexistence line}
\author{Ronald Benjamin and J{\"u}rgen Horbach}
\affiliation{Institut f{\"u}r Theoretische Physik II, Universit{\"a}t D{\"u}sseldorf,
Universit\"atsstra\ss e 1, 40225 D{\"u}sseldorf, Germany}

\begin{abstract}
Kinetics of crystal-growth is investigated along the solid-liquid
coexistence line for the (100), (110) and (111) orientations of the
Lennard-Jones and Weeks-Chandler-Andersen fcc crystal-liquid interface,
using non-equilibrium molecular dynamics simulations. A slowing down
of the growth kinetics along the coexistence line is observed, which is
mostly a temperature effect, with other quantities such as the melting
pressure and liquid self-diffusion coefficient having a negligible impact.
The growth kinetics of the two potentials become similar at large values
of the melting temperature and pressure, when both resemble a purely
repulsive soft-sphere potential.  Classical models of crystallization from
the melt are in reasonable qualitative agreement with our simulation data.
Finally, several one-phase empirical melting/freezing rules are studied
with respect to their validity along the coexistence line.
\end{abstract}

\maketitle

\section{Introduction}
\label{sec:intro}
A central quantity for the understanding of crystallization processes
from the undercooled liquid is the kinetic growth coefficient
\cite{palberg2014}.  The kinetic growth coefficient, $\mu$, is defined
as the constant of proportionality between the velocity $v_{\rm i}$ with
which the crystal-liquid interface moves and the interfacial undercooling,
$\Delta T = T_{\rm M} - T$,
\begin{equation}
v_{\rm i}=\mu \Delta T,
\label{eq_vi}
\end{equation}
with $T_{\rm M}$ the melting temperature. Note that Eq.~(\ref{eq_vi}) is
only expected to hold if the undercooling $\Delta T$ is sufficiently
small.  Magnitude and anisotropy of the kinetic growth coefficient
play a dominant role in determining the morphology of the growing
crystal~\cite{bragard2002,langer1980} and are also essential
parameters required for the continuum modelling of solidification
processes~\cite{bragard2002}.

Experimental measurements of the kinetic growth coefficient have
been scarce, with the exception of a few studies on metallic
systems~\cite{willnecker_herlach89,rodway91} and white phosphorous
(P$_4$)~\cite{glicksman67}. However, atomistic simulation techniques
\cite{allen-tildesley87,binder2004} such as Molecular Dynamics (MD)
provide detailed information about the microscopic structure and
dynamics of the interface region. Thus, atomistic simulations can be
used to test various analytical approaches to describe crystal growth
such as the Wilson-Frenkel~\cite{wilson1900,frenkel1932}
and Broughton-Gilmer-Jackson~\cite{jackson1982,jackson2002} model.

Different simulation techniques have been
developed for the investigation of crystal growth
kinetics~\cite{jackson1982,jackson2002,broughton88,celestini2002,monk2002,hoyt-asta-karma2002,hoyt-asta-karma2003,hoyt-asta2002,amini2006,benet2014,turci-schilling2014,tepper-briels1997,briels2001,briels2002}.
In the capillary fluctuation
method~\cite{monk2002,hoyt-asta-karma2002,hoyt-asta2002,hoyt-asta-karma2003,amini2006,benet2014,turci-schilling2014},
the kinetic growth coefficient $\mu$ is obtained from equilibrium MD
simulations by analyzing the height fluctuations of the crystal-liquid
interface at coexistence.  Using this approach, $\mu$ has been computed
for hard spheres~\cite{amini2006}, metals~\cite{hoyt-asta2002}, a
Lennard-Jones system and the TIP4P/2005 water model~\cite{benet2014}. A
variation of the capillary fluctuation method has been proposed by
Tepper and Briels~\cite{tepper-briels1997,briels2001,briels2002}. In
their approach, $\mu$ is extracted from the equilibrium fluctuations of
the number of crystalline atoms in an inhomogeneous solid-liquid system.

Another widely used approach to obtain kinetic growth coefficients is the
free solidification method (FSM)~\cite{actamaterhoyt1999,hoyt-asta2002}
which is based on non-equilibrium MD. Here, one simulates inhomogeneous
systems where the crystal is separated from the liquid phase via
two planar interfaces (two interfaces appear due to periodic boundary
conditions). By monitoring the rate of change of the system's volume with
respect to time, the kinetic growth coefficient, $\mu$, can be determined.
This approach has been applied to various one- and two-component
metals~\cite{hoyt2004,hoyt-asta2002,kuhn2013,kerrache2008,xia2007,sunastahoyt2004,tymczak1990,gao2010,timan2009,timan2010}
as well as Lennard-Jones systems~\cite{huitema99,briels2001}. The estimated 
values of $\mu$ for metallic systems, obtained from FSM, have been shown 
to be in good agreement with those obtained for hard spheres from 
the capillary fluctuation method \cite{amini2006}.

For systems with a crystal face-centered cubic (fcc) phase, it has been
suggested from the latter simulation studies that the diffusion-limited
classical Wilson-Frenkel model of crystal growth has to be modified into
a collision-limited growth model to explain the high crystal growth rates
corresponding to the (100) orientation of the crystal-liquid interface.
Moreover, the collision-limited growth model seems to be a good predictor
of $\mu$ for the (110) interface of metallic systems~\cite{hoyt-asta2002},
too. Only the (111) interface tends to follow the Wilson-Frenkel
kinetics~\cite{broughton88}.

Most of the above studies have been done under ``ambient conditions'',
i.e.~at the melting temperature corresponding to zero pressure conditions.
Little is known about the dependence of the kinetic growth coefficient
on pressure and temperature along the coexistence line.  In this work,
we investigate the growth kinetics of the (100), (110), and (111)
orientations of a fcc crystal-liquid interface for two different models,
employing the FSM: (i) a force-shifted Lennard-Jones (fsLJ) potential and
(ii) a purely repulsive Weeks-Chandler-Andersen (WCA) potential. For
both models, systems at various pressures and temperatures along the
coexistence line are studied. From the FSM simulations, the coexisting
temperatures and pressures as well as the kinetic growth coefficients
are obtained.

Our results indicate that the growth kinetics depends only weakly on
pressure while it is significantly affected by a change of the melting
temperature. For the fsLJ potential, there is an initial regime where the
coexistence pressure changes by two orders of magnitude while the melting
temperature remains essentially unchanged. In this regime, the kinetic
growth coefficients are almost constant as a function of pressure. For both
the fsLJ and the WCA models, however, an increasing melting temperature
$T_{\rm M}$ (with increasing the ``melting'' pressure $P_{\rm M}$) is
associated with a slowing-down of the growth kinetics. At high values of
$T_{\rm M}$ and $P_{\rm M}$, the kinetic growth coefficients in reduced
units tend to reach asymptotic values which correspond to the ones found
for hard spheres, though not identical. In our analysis, we discuss to
what extent the collision-limited crystal growth model of Broughton,
Gilmer, and Jackson (BGJ) is valid and relate the growth kinetics to the
self-diffusion coefficient of the liquid, the entropy of fusion and the
liquid and crystal coexistence densities. In this context, we also discuss
empirical rules of melting and freezing~\cite{loewen1994,monson2000},
namely the Lindemann~\cite{lindemann1910}, the Ravech\'{e}, Mountain and 
Streett~\cite{rms1974} and the Hansen-Verlet~\cite{hv6970} criterion.

In the next Section, we describe the two interaction potentials considered
in this work. In Section III, we outline the FSM, followed by Section
IV on the simulation details. The results are presented in Section V,
and finally we come to the conclusions in Section VI.

\section{Interaction Potentials}
\label{sec:potential}
%

%%%
\begin{figure}
\includegraphics[width=3.0in]{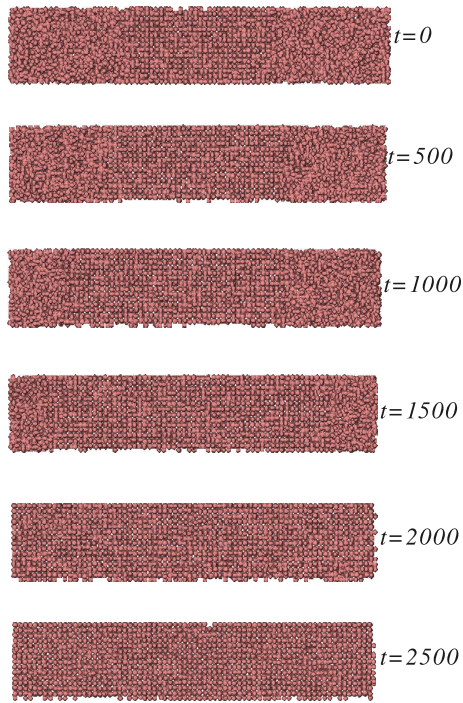}
\caption{(Color online) Time evolution of a crystal-liquid interface
during growth ($T<T_{\rm M}$), corresponding to the (100) orientation
of the fsLJ system at the pressure $P=1.0$ and temperature $T=0.65$,
represented by snapshots of the system at various times $t$ (the 
melting temperature is $T_{\rm M}=0.653$). Initially (topmost
configuration), there is a crystal sandwiched by equal amounts of
liquid on both sides. With time the liquid portion shrinks gradually
and eventually the entire system crystallizes. Time $t$ is measured in
units of $\tau$ (see text). \label{fig1}}
\end{figure}
%%%

Simulations are carried our for two different models that are derived from
the Lennard-Jones (LJ) potential. This potential describes the interaction
between two particles separated by a distance $r$ by the function
\begin{equation}
\phi(r) = 4 \varepsilon \left[ 
\left(\frac{\sigma}{r} \right)^{12} - \left(\frac{\sigma}{r} \right)^{6}
\right] \, ,
\end{equation}
with $\varepsilon$ and $\sigma$ being two parameters setting respectively
the microscopic energy and length scales for two neighboring particles.

The first model considered in this work is a force-shifted Lennard-Jones
(fsLJ) model, defined by
\begin{equation}
U_{\text{fsLJ}}(r) = \left\{
\begin{array}{lr}
\phi(r)-\phi(r_{c})-\phi^{\prime}(r_{c})[r-r_{c}]
& r<r_{\rm c} \\
0 & {\rm otherwise}
\end{array}
\right.
\end{equation}
with $\phi^{\prime} \equiv d\phi/dr$. The cut-off of the potential is
set to $r_{\rm c}=2.5\,\sigma$.

The second model is a Weeks-Chandler-Andersen (WCA) potential, i.e.~a
LJ potential which is cut off at its minimum at $r_{\rm c}=2^{1/6}\,\sigma$
and shifted to zero,
\begin{equation}
U_{\text{WCA}}(r) = \left\{
\begin{array}{lr}
\phi(r)+ \varepsilon & \quad r<2^{1/6}\,\sigma \\
0 & \quad {\rm otherwise} \, .
\end{array}
\right.
\end{equation}
Thus, the WCA model is a purely repulsive potential. In the fsLJ model,
only at very high coexistence pressures the repulsive part is expected
to dominate kinetic properties and phase behavior and so we can study
how attractions between the particles affect crystal growth along the
coexistence line going from low to high coexistence pressures.

In the following, energies and lengths are expressed in units of
$\varepsilon$ and $\sigma$, respectively, and the masses of the particles
are set to $m=1$. Thus, thermal energy, $k_{\rm B}T$ (with the Boltzmann
constant $k_{\rm B}=1$ and $T$ the temperature), and pressure, $P$, are
expressed in units of $\varepsilon$ and $\varepsilon/\sigma^{\rm 3}$,
respectively.  Time is measured in units of $\tau = \sqrt{m\sigma^{\rm
2}/\varepsilon}$, while the kinetic growth coefficient is reported in
units of $k_{\rm B}/\sqrt{m \varepsilon}$.

\section{Free solidification method (FSM)}
\label{sec:fsm}
A crystal in contact with its melt at a temperature $T$ below
(above) the melting temperature $T_{\rm M}$ (cf.~Fig.~\ref{fig1}) will
grow (shrink) until the entire system crystallizes (melts).  The FSM
\cite{actamaterhoyt1999,hoyt-asta2002,kerrache2008,timan2009,timan2010}
is an approach to compute the crystal growth (or melting) rate as well
as the coexistence temperature $T_{\rm M}$ at a given pressure $P$. The
starting point are standard isothermal-isobaric MD simulations at various
temperatures $T$ and at a particular value of the pressure, $P$, keeping
the number of particles $N$ constant.  From these $NPT$ simulations,
the temperature dependence of the density of the crystal (in our case
a fcc crystal) as well as the melt are determined. Gradually increasing
the temperature of the crystal leads to melting at a temperature $T_1$
while the subsequent gradual reduction of the temperature leads to
re-crystallization at a different temperature $T_2$.  Thus, hysteresis is
observed, i.e.~the heating and cooling curves do not follow the same path.
Such ``heating-cooling'' plots indicate the region in which the melting
temperature, $T_{\rm M}$, is located.

\begin{figure}
\includegraphics[width=3.0in]{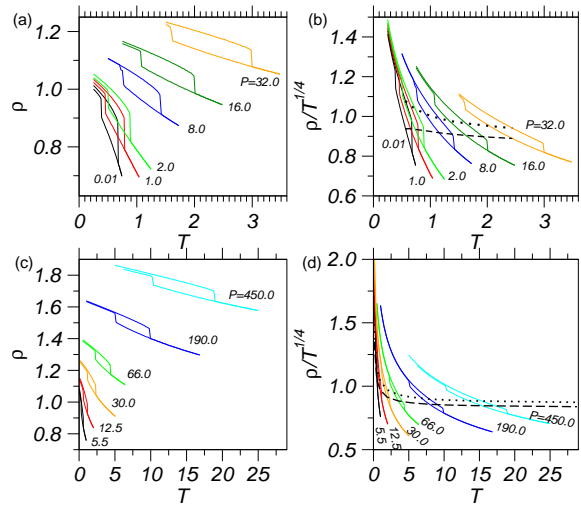}
\caption{\label{fig2}(Color online) Heating-cooling curves in the
density-temperature plane corresponding to (a) the fsLJ and (c) the WCA
model at various applied pressures. Also plotted is the temperature
dependence of $\rho/T^{\rm 1/4}$ for the fsLJ and the WCA model in panels
(b) and (d), respectively. The dotted and dashed lines in panels (b)
and (d) represent the coexistence values of the crystal and liquid,
respectively.}
\end{figure}

Now, the FSM scheme consists of the following
steps~\cite{kerrache2008,timan2009,timan2010}: First, $N$ atoms are
arranged on a fcc lattice in an elongated simulation box of size
$L_{x}\times L_{y} \times L_{z}$ with the desired orientation of the
crystal pointing in $z$ direction and the length of the system along
the $z$ direction being approximately five times that in $x$ and $y$
directions (cf.~Fig.~\ref{fig1}). At each temperature and pressure,
the density of the fcc crystal is obtained from the aforementioned
heating-cooling plots (Fig.~\ref{fig2}). Then, at a given temperature
and pressure the system is equilibrated in a $NP_{x}P_{y}P_{z}T$
ensemble~\cite{sunastahoyt2004} (with $P_{x}$, $P_{y}$, and $P_{z}$
respectively the pressures along the $x$, $y$ and $z$ directions)
in the range in which hysteresis is observed. The reason for carrying
out simulations in the constant pressure ensemble is to ensure that
the crystal is free of any residual stress along the three Cartesian
axes.  Moreover, for maintaining constant pressure along the different
Cartesian axes, simulations are carried out in the  $NP_{x}P_{y}P_{z}T$
ensemble rather than in the $NPT$ ensemble since the simulation box is a
cuboid with unequal lengths along the  different Cartesian axes. After
equilibration is reached, the average length of the simulation box in
the $x$ and $y$ directions are determined.

After relaxing the crystal sample in the first step, two-thirds of the
atoms in the middle of the box are fixed and the rest of the system
is heated up to a high temperature, $T>>T_{\rm M}$, to eventually melt
it. In this step, the lengths of the simulation box in the $x$ and $y$
directions are fixed to the average lengths obtained from the previous
$NP_{x}P_{y}P_{z}T$ ensemble run, and the simulation is carried out in
the $NP_{z}AT$ ensemble by varying only the length $L_{\rm z}$ (i.e.,
maintaining constant pressure $P_{z}$). Here, $A$ corresponds to the area
of the system ($A=L_{\rm x}\times L_{\rm y}$). In the third step, the
temperature of the whole system is set back to the initial temperature in
which the crystal was prepared, with the atoms in the middle of the region
still fixed. This simulation in the $NP_zAT$ ensemble runs for a short period,
just long enough to cool the melted region to the desired temperature.
Finally, all the particles are allowed to move and the simulation in
the $NP_zAT$ ensemble is continued.

In the steady-state, the length of the system, $L_z$, varies linearly
with time and one obtains the change of $L_z$ per unit time, $\dot{L}_z$,
from the slope of $L_z(t)$.  From $\dot{L}_z$ and a mass balance equation,
it is straightforward to obtain the interface velocity via \cite{huitema99}
\begin{equation}
v_{\rm i}= - \frac{{\rho}_{\rm l}\dot{L}_z}{2(\rho_{\rm c}-\rho_{\rm l})}
\label{eq_vi2}
\end{equation}
with $\rho_{\rm c}$ and $\rho_{\rm l}$ the bulk densities of the solid
and liquid phase, respectively.  Close to the melting temperature the
interface velocity varies linearly with temperature.  From a linear
fit to the temperature dependence of the interface velocity, the kinetic
growth coefficient $\mu$ as well as the melting temperature $T_{\rm M}$
are obtained, cf.~Eq.~(\ref{eq_vi}).

%%%
%%%

%
\section{Simulation Details}
\label{sec:simul-detail}
To integrate the equations of motion, the velocity Verlet
algorithm \cite{allen-tildesley87} is used. Pressure is
kept constant by coupling the system to an Andersen barostat
\cite{allen-tildesley87,andersen80,sunastahoyt2004}. The coupling
parameter of the barostat (the mass of the piston) is set to $M=0.001$
for the $NPT$ runs to determine the ``heating-cooling'' curves and to
$M=10$ for all the FSM simulations in the $NP_{\rm x}P_{\rm y}P_{\rm z}T$
and the $NP_{\rm z}AT$ ensemble.  To keep the temperature constant, the
system is coupled to a stochastic heat bath by assigning every 200 time
steps random velocities to the atoms sampled from a Maxwell-Boltzmann
distribution. The reduced time step is taken to be $\delta t =0.005\tau$
for all the simulations.

To compute the density of the crystalline and liquid phases at
coexistence and to determine the density-temperature hysteresis curves,
$NPT$ simulations with $N=2048$ particles are carried out at various
values of the pressure in the interval $0.005 \le P \le 32$ for the fsLJ
potential and in the interval $1.0 \le P \le 1020$ for the WCA potential.
Initially, the particles are placed on a fcc lattice in a cubic cell of
dimensions $L\times L\times L$. We equilibrate the system for 25000 time
steps and then perform production runs for another 25000 steps. From the
data collected during the production runs, the volume of the system is
determined to obtain the equilibrium density. The temperature of the
system is raised by a small step and the above procedure is repeated
to obtain the density for the next higher temperature. This process is
continued until the crystal melts to form the liquid phase. The same
procedure is followed to obtain the bulk liquid equilibrium density
by gradually lowering the temperature of the system in small steps and
calculating the equilibrium density at each temperature.

For the FSM simulations of the fsLJ model, particles are placed in an
elongated box of size $L\times L\times 5L$. Systems containing $N=34560$,
$35088$ and $35700$ particles are considered respectively for the
(100), (110), and (111) orientation of the fcc lattice in $z$ direction.
With the same relative lengths of the simulation box along the $x$, $y$
and $z$ directions, simulations with  $N=14580$, $15444$ and $15795$
particles are carried out for the WCA model along the (100), (110),
and (111) orientation, respectively.  At several temperatures in the
hysteresis region (see Sec.~\ref{sec:fsm}) crystals are  
equilibrated for $75000$ time-steps in the $NP_{\rm x}P_{\rm y}P_{\rm z}T$ 
ensemble.  From the last $10000$ time-steps, the average lengths of 
the simulation cell along the $x$ and $y$ directions is determined. 
$L_{\rm x}$ and $L_{\rm y}$ are fixed to this value to carry out 
simulations for the next step in the $NP_{\rm z}AT$ ensemble.

In the second step, the crystalline particles in the middle one-third
of the box are fixed, while particles in the remaining region are
equilibrated at a high temperature for $150000$ time steps to melt
the system.  Then, keeping the crystalline region fixed, the liquid
is cooled to the desired temperature in a short run of $10000$ time
steps. Finally, all the particles are allowed to move, performing a
run over $500000$ time steps from which the interface velocity $v_{\rm
i}$ is determined. Statistical errors were determined from $10-20$
independent realizations.

\section{Results}
\label{sec:results}
\subsection{Heating-cooling plots}
Figures \ref{fig2}a and c show heating-cooling plots of the
density at different pressures for the fsLJ and the WCA model,
respectively.  With increasing pressure, hysteresis is observed
in a larger temperature interval.  As we shall see below (see
Sect.~\ref{subsec:crystalgrowthkinetics}), this is associated with
a gradual slowing down of crystal growth kinetics.  By scaling the
density by a factor $1/T^{\rm 1/4}$, all the curves corresponding
to the various pressures tend to fall in a similar range along the
$y$-axis (see Fig.~\ref{fig2}b and d). Also, the coexistence values
of the quantity $\rho/T^{\rm 1/4}$ for the crystal and liquid
asymptotically reach a similar constant value for both the fsLJ and
WCA model, indicating that the phase behavior of both systems at high
temperature and pressure is similar. We discuss this point in more detail
in Sect.~\ref{subsec:coexistence}.

%%%
\begin{figure}
\includegraphics[width=3.0in]{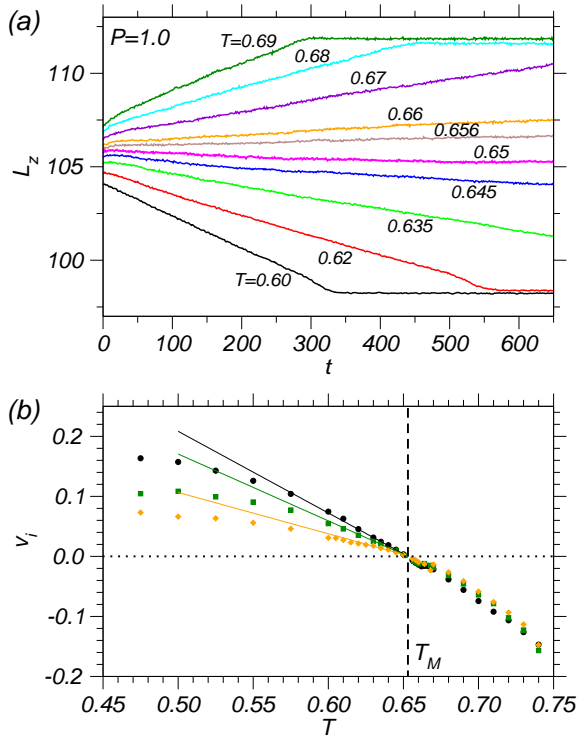}
\caption{(Color online) (a) Length of the system along the $z$
direction, $L_{\rm z}$ vs.~time for different temperatures at $P=1.0$. 
(b) Interfacial velocity as a function of the temperature at $P=1.0$
for the (100) (circles), (110) (squares), and (111) (diamonds) crystal 
orientations. Straight lines are linear fits to the data at low
undercoolings (see text).\label{fig3}}
\end{figure}
%%%

%
\subsection{Interface velocity}
\label{subsec:interf_vel}
%

%%%
\begin{figure}
\includegraphics[width=3.0in]{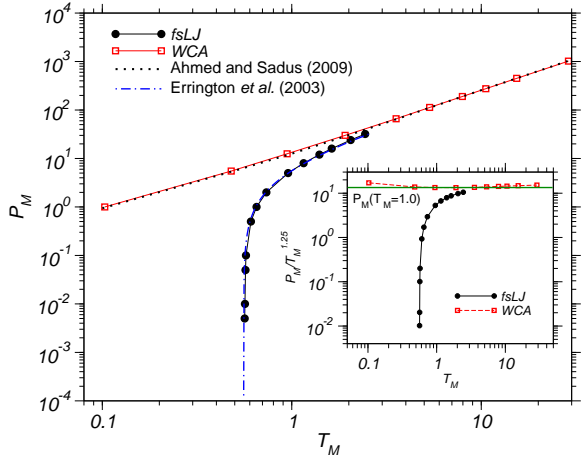}
\caption{(Color online) Phase diagram in the pressure-temperature plane
for the fsLJ and the WCA model, as indicated.The dotted and dash-dotted
lines represent coexistence values obtained by Ahmed and Sadus for the WCA 
potential~\cite{sadus2009} and by Errington {\it et al.} for the fsLJ
potential~\cite{torquato2003}. The inset shows the quantity
$P_{\rm M}/{T_{\rm M}}^{1.25}$ as a function of the melting temperature
$T_{\rm M}$. The green solid line corresponds to the melting pressure
for the WCA potential at $T_{\rm M}=1.0$. \label{fig4}}
\end{figure}
%%

%%%
\begin{figure}
\includegraphics[width=3.0in]{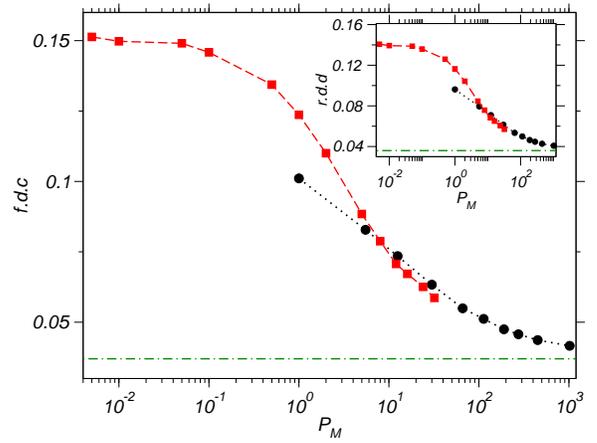}
\caption{(Color online) Variation of f.d.c.~and r.d.d.~(inset) as a
function of the coexistence pressure for the fsLJ and WCA potential. The
green dashed horizontal lines represent the corresponding values for
the inverse twelfth-power potential (see text). \label{fig5}}
\end{figure}
%%

%%%
\begin{figure}
\includegraphics[width=3.0in]{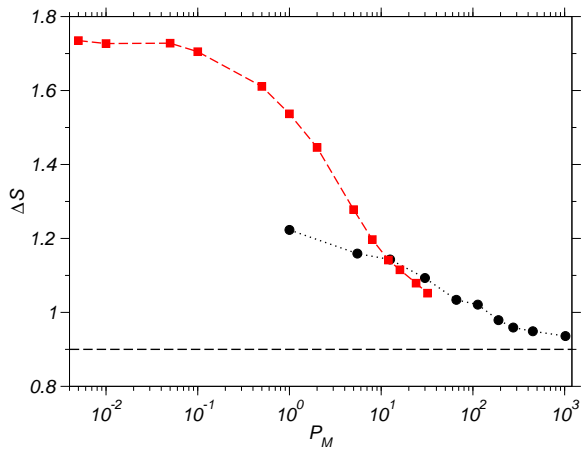}
\caption{(Color online) Entropy of fusion as a function of the melting
pressure corresponding to the WCA and the fsLJ system. The green dashed
horizontal line represents $\Delta S$ for the inverse twelfth-power
potential (see text). \label{fig6}}
\end{figure}

The behavior of the crystal-melt interface depends on the temperature
at which it is simulated. The crystal will grow below the melting
temperature while above it, melting will occur. Since the crystal
density is higher than the melt, the length of the system along which
the interface is oriented will increase during melting and shrink
during crystallization. In Fig.~\ref{fig3}a, we plot the length of the
system along the $z$ direction, $L_{\rm z}$, versus time for different
temperatures corresponding to the fsLJ potential at $P=1.0$. The data
is averaged over $10$ independent realizations.  
Figure~\ref{fig3}a  shows that after a transient period at the
beginning, $L_{\rm z}$ varies linearly with time when the steady state
is reached and ultimately reaches a constant at long times, when the
whole system has either crystallized or melted.  Just prior to this,
we find a non-linear regime where the crystallization and melting are
much faster than in the linear steady-state regime, because one of the
phases has shrunk to such a small size that the two interfaces interact
with each other (see the two bottom-most curves in Fig.~\ref{fig3}a).

Figure~\ref{fig3}b shows $v_{\rm i}$ as a function of temperature for the
three different crystal orientations i.e.~$(100)$, $(110)$ and $(111)$
at $P=1.0$.  The melting temperature is dictated by thermodynamics
and is expected to be identical for all crystal orientations. At small
undercoolings, the system is in the linear response regime. Hence,
around $T_{M}$,  the simulation data for $v_{\rm i}$ can be fitted by a
linear law $v_{\rm i}=\mu (T_{M}-T)$, with $v_{\rm i}=0$ at $T_{\rm M}$.
From Fig.~\ref{fig3}b, we find that for the (100) crystal orientation,
$v_{\rm i}$ vanishes around $T=0.653$. For the (110) and (111)
crystal orientations, $v_{\rm i}$ approaches zero respectively 
at $T=0.653$ and $T=0.655$.

\subsection{Thermodynamic properties at coexistence}
\label{subsec:coexistence}
The FSM simulations were carried out at different pressures
to obtain the respective melting temperatures. Figure~\ref{fig4} shows
the phase diagram of the fsLJ system and the WCA potential along the
$P-T$ plane. Our simulation data corresponding to the coexistence
conditions is in very good agreement with that obtained previously
using Gibbs-Duhem integration from a known melting temperature and
pressure at a single coexistence point for the fsLJ~\cite{torquato2003}
and the WCA model~\cite{sadus2009}. At low pressures, there is a significant
difference between the coexistence curves of the two potentials on account
of the different roles played by the $-1/r^{\rm 6}$ term.  Due to this
attractive part of the fsLJ potential, the particles sit at the potential
well at low pressures ($P_{\rm M} \lesssim 0.5$) and as a result the
melting temperature stays almost the same even when the potential changes
by two orders of magnitude (from $P=0.005$ to $P=0.5$).  For $P_{\rm
M}>0.5$, the melting temperature increases rapidly with the pressure
as the repulsive part of the potential becomes dominant while the
attractive term plays less and less of a role in determining the phase
behavior. One can clearly identify these two regimes in the coexistence
line corresponding to the fsLJ potential as shown in Fig.~\ref{fig4}.

Figure \ref{fig4} also shows that the $P-T$ coexistence curve of the
WCA model is almost a straight continuous line in the whole considered
range of melting temperatures. Moreover, at high values of $T_{\rm M}$
the coexistence line of the fsLJ model seems to become identical to
that of the WCA model.  This behavior is expected because at high
temperatures the phase behavior of both models is dominated by the
repulsive $1/r^{12}$ interactions, in agreement with the findings in
Refs.~\cite{hoover1971,sadus2009,ahmad-sadus2009}.

It is interesting, therefore, to compare the coexistence behavior
of the fsLJ and the WCA potentials with the inverse 12th-power
soft-sphere potential, $U(r)=\varepsilon (\sigma/r)^{n}$
($n=12$).  Inverse-power law potentials are fully determined by one
parameter (here $\varepsilon \sigma^{\rm n}$) and coexistence is
fully specified by a single quantity, $\Gamma_{n}=\rho {T_{\rm
M}}^{-3/n}$~\cite{hansen-liquid-theory2006,laird-davidchack2005}.  As a
consequence, the reduced melting pressure shows a power-law scaling
with respect to the reduced melting temperature, $P_{\rm M}=P_{\rm
1}{T_{\rm M}}^{1+3/n}$.  For $n=12$, this relation reduces to $P_{\rm
M}=P_{\rm 1}{T_{\rm M}}^{\rm 1.25}$, where $P_{\rm 1}$ is the pressure
corresponding to $T_{\rm M}=1.0$.  In the inset of Fig.~\ref{fig4},
we plot the quantity $P_{\rm M}/{T_{\rm M}}^{\rm 1.25}$ as a function
of the melting temperature for both the WCA and the fsLJ model. In case
of the WCA potential ($P_{\rm 1}=13.41$), this relation is satisfied for
$T_{\rm M}>1.0$. For the fsLJ potential, at larger melting temperatures,
this quantity tends to the same value as in case of the WCA potential
i.e.~$P_{\rm M}/T_{\rm M}^{\rm 1.25}=13.41$. However, the melting
temperature corresponding to this value of the pressure is around $T_{\rm
M}=1.48$, indicating that for $T_{\rm M}>1.48$ (or $P_{\rm M}> 14.0$)
the phase behavior of the fsLJ potential approaches that of the purely
repulsive inverse 12th-power soft-sphere potential.

In Fig.~\ref{fig2}b and d, we have plotted the scaled density $\rho
{T_{\rm M}}^{-1/4}$ as a function of temperature that tends to approach
a constant value in the high-temperature limit that corresponds to
that of a $1/r^{12}$ soft-sphere potential.  For the latter potential,
the estimate of Hoover {\it et al.}~\cite{hoover1971} for the crystal
(liquid) coexistence value of $\rho {T_{\rm M}}^{-1/4}$ is $0.844$
($0.813$), while for the fsLJ and WCA interaction potentials the
coexistence values of the crystal (liquid) at the highest pressures
considered are $0.941$ ($0.889$) and $0.873$ ($0.838$) at $P_{\rm M}=32$
and $1020$, respectively.  Thus, also with respect to the scaled density
$\rho T_{\rm M}^{-1/4}$ both the fsLJ and the WCA model approach the value
obtained for the $r^{-12}$ soft-sphere potential. 

Along the coexistence line, we now discuss further thermodynamic
properties, namely the fractional density change at freezing,
${\rm f.d.c.}=(\rho_{\rm c}-\rho_{\rm l})/\rho_{\rm l}$, also known as
the miscibility gap, the relative density difference at freezing,
${\rm r.d.d.}=2(\rho_{\rm c}-\rho_{\rm l})/(\rho_{\rm c}+\rho_{\rm l})$, 
and the entropy of fusion $\Delta S =l/T_{\rm M}$.  Here,  $l$ is 
the latent heat as obtained from the difference of the enthalpies 
of liquid and crystal at coexistence.

In Fig.~\ref{fig5}, we show the ratios f.d.c.~and r.d.d.~(inset) as a
function of the melting pressure for the two interaction potentials.
Both the miscibility gap and the relative density difference at freezing
decrease along the coexistence line. Data corresponding to the WCA
potential are in good qualitative agreement with those obtained in an
earlier work~\cite{sadus2009} but the magnitudes of the f.d.c.~and
r.d.d.~reported by us are slightly higher. At larger pressures the
fsLJ data tends to attain similar values as those of the WCA potential
though somewhat lower. The f.d.c.~and r.d.d.~ratios corresponding to
the $r^{-12}$ potential are $0.038$ and $0.037$~\cite{hoover1971},
respectively (indicated by the horizontal lines in Fig.~\ref{fig5}).
Figure~\ref{fig5} clearly shows that at large pressures both the WCA and
fsLJ data converge to the same values as those of the $r^{-12}$ potential.

The entropy of fusion (see Fig.~\ref{fig6}) decreases with increasing
temperature and pressure for both interaction potentials in qualitative
agreement with a previous work~\cite{sadus2009}. This indicates that
at lower melting temperatures there is a greater positional ordering
of the solid phase as compared to that at higher temperatures.  At the
lower pressures, the entropy of fusion for the fsLJ potential approaches
a constant value as the melting temperature changes little. At larger
pressures, the values of $\Delta S$ converge to that of the $r^{-12}$
potential, $\Delta S=0.9$~\cite{hoover1971}.

We have seen that with respect to the thermodynamic properties, the fsLJ
and the WCA model become similar in the high-temperature limit where,
in both cases, repulsive $r^{-12}$ interactions dominate the phase
behavior. Below, we show (Sect.~\ref{subsec:crystalgrowthkinetics})
that the crystal growth kinetics of the fsLJ and the WCA model
also becomes similar at high temperatures/pressures.

%%%
\begin{figure}
\includegraphics[width=3.0in]{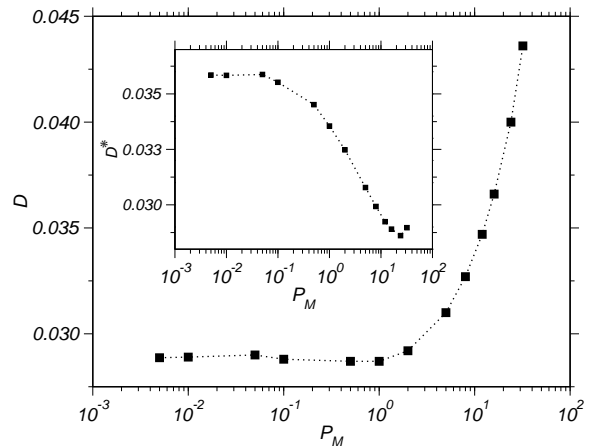}
\caption{(Color online) The self-diffusion coefficients $D$ (main figure)
and $D^{\rm s\ast}$ (in the inset) as a function of the coexistence
pressure $P_{\rm M}$ corresponding to the fsLJ potential. \label{fig7}}
\end{figure}
%%

%%%
\begin{figure}
\includegraphics[width=3.0in]{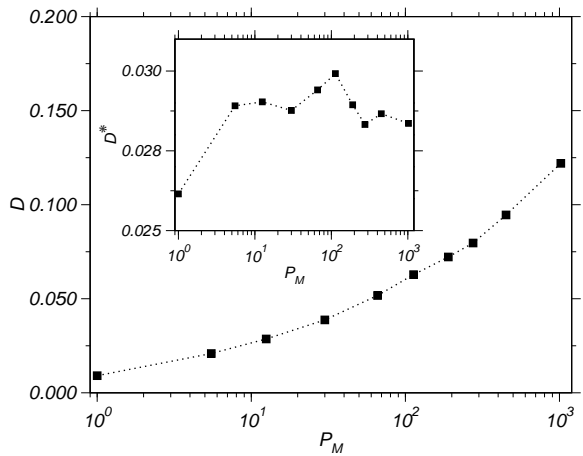}
\caption{(Color online) The self-diffusion coefficients $D$ (main figure)
and $D^{\rm \ast}$ (in the inset) as a function of the coexistence
pressure $P_{\rm M}$ corresponding to the WCA potential. \label{fig8}}
\end{figure}
\subsection{Self-diffusion coefficient of the liquid}
One of the classical models of crystal growth kinetics is the
Wilson-Frenkel model~\cite{wilson1900,frenkel1932} which describes crystal
growth by an activated process, limited by the self-diffusion of the
atoms in the liquid phase. Thus, this model predicts that the diffusion
dynamics of the liquid strongly affects the growth kinetics. Therefore,
we now analyze the self-diffusion coefficient of the liquid phase for
temperatures and pressures along the coexistence line.

The self-diffusion coefficient, $D$ is computed from the mean-squared
displacement of a tagged particle \cite{binder2004}, $\delta r^2(t)
= \langle (\vec{r}_{\rm tag}(t) - \vec{r}_{\rm tag}(0))^2 \rangle$
(with $\vec{r}_{\rm tag}(t)$ the position of the tagged particle at time
$t$), via the Einstein relation, $D = \lim_{\rm t\rightarrow \infty}
\delta r^2(t)/6t$.

In Figs.~\ref{fig7} and ~\ref{fig8}, we display $D$ along the $P-T$
coexistence line for the fsLJ and the WCA model, respectively.  As is
evident from Fig.~\ref{fig7}, $D$ for the fsLJ model remains unchanged
in the pressure range $0.005 \leq P_{\rm M} \leq 0.5$, where the
corresponding melting temperature changes very little.  However, when the
melting temperature significantly increases by about a factor of three the
diffusion coefficient also increases by about 50\%.  For the WCA potential
(Fig.~\ref{fig8}), the diffusion constant increases with increasing
pressure due to the change in $T_{\rm M}$ all along the coexistence
line. Overall, the increase in diffusion coefficient with increasing
melting pressures (provided $T_{\rm M}$ increases), might indicate that the
growth kinetics become faster along the coexistence line. However, as
reported below in Sect.~\ref{subsec:crystalgrowthkinetics}, the kinetic
growth coefficient decreases at high pressures, indicating that the
self-diffusion coefficient of the liquid does not play a significant
role in determining the growth kinetics.

For comparing the fsLJ and WCA potentials, the scaled diffusion constant,
$D^{\rm \ast}=D/D_{\rm sc}$ (with $D_{\rm sc}= \sqrt{k_{\rm B}T_{\rm
M}/m}/{\rho_{\rm l}}^{1/3}$ being a natural MD scale), is shown in the
insets of Figs.~\ref{fig7} and \ref{fig8}. $D^{\rm \ast}$ for the fsLJ
potential decreases with increasing pressure and approaches a value
between $0.029$ and $0.030$ at large pressures. In case of the WCA
potential, $D^{\rm \ast}$ slightly increases at low pressures and then,
in the pressure range $5.5 \leq P \leq 1020$, remains almost unchanged,
also at a value between $0.029$ and $0.030$.  Thus, also with respect to
the scaled self-diffusion coefficient $D^{\rm \ast}$ similar values are
obtained for the fsLJ and the WCA model at high temperatures/pressures.

%%% 
\begin{figure}[htbp] 
\includegraphics[width=3.0in]{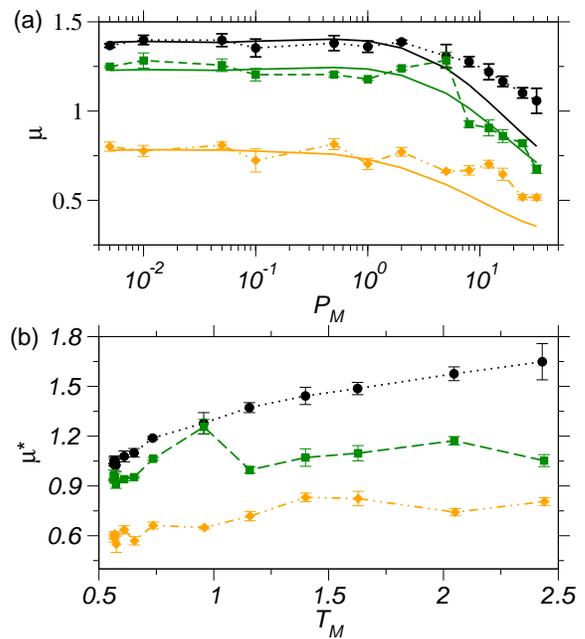}
\caption{(Color online) The kinetic growth coefficient $\mu$
corresponding to the fsLJ potential for the three crystal orientations
(100) (circles), (110) (squares), and (111) (diamonds) (a) as a function
of the coexistence pressure $P_{\rm M}$ and (b) as a function of the
coexistence temperature. In (a), the solid lines  represent the fits
obtained from the classical models. In (b), $\mu^\star=\mu \sqrt{T_{\rm M}}$
is shown. \label{fig9}} 
\end{figure} 
%%%

%
\subsection{Crystal growth kinetics}
\label{subsec:crystalgrowthkinetics}
\begin{figure}[htbp]
\includegraphics[width=3.0in]{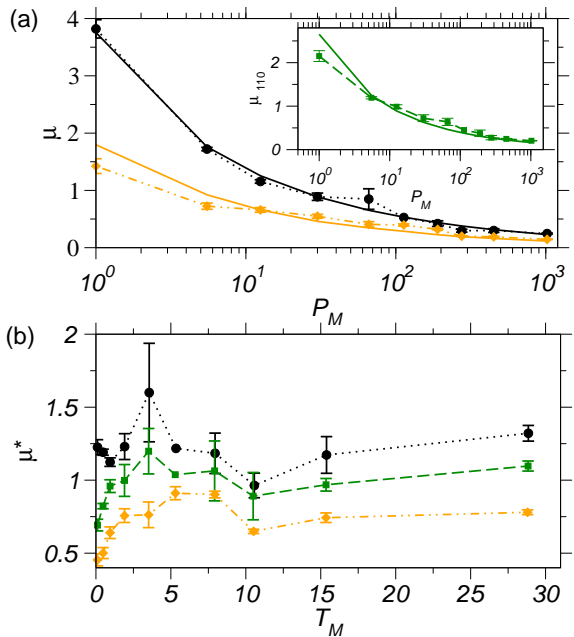}
\caption{(Color online) The kinetic growth coefficient $\mu$ corresponding
to the WCA potential for the three crystal orientations (100), (110),
and (111), (a) as a function of the coexistence pressure $P_{\rm M}$,
and (b) as a function of the coexistence temperature $T_{\rm M}$. Note that in
(b), $\mu^\star = \mu \sqrt{T_{\rm M}}$ is shown. For clarity, in
(a), $\mu$ for the (110) orientation is shown in the inset. Representation 
of symbols is same as in Fig.~\ref{fig9}.\label{fig10}}
\end{figure}

The kinetic growth coefficient $\mu$ is extracted from the slope of the
linear fit to the interfacial velocity $v_{\rm i}$ at small undercoolings.
Our results for both the fsLJ and WCA model confirm observations of prior
studies~\cite{amini2006,hoyt-asta2002} regarding the magnitude of $\mu$
for the different orientations: $\mu_{\rm 100}>\mu_{\rm 110}>\mu_{\rm
111}$. The data for the three different crystal orientations along the
coexistence line are reported in Figs.~\ref{fig9}a and \ref{fig10}a
for the fsLJ and WCA model, respectively. For the fsLJ model, $\mu$
remains essentially constant in the pressure range $0.005 \le P \le
1.0$ where the melting temperature changes only weakly. For larger
pressures, when the melting temperature increases rather sharply,
the kinetic growth coefficient decreases, indicating this decrease to
be of a thermal origin. Similarly, in case of the WCA potential, $\mu$
decreases with increasing pressure reflecting the variation of the melting
temperature with respect to the coexistence pressure (Fig.~\ref{fig10}a).
At a pressure of $P_{\rm M}\approx 30$, the values of $\mu$ for both
models are already in good quantitative agreement.

%%%
%%%

To compare the values of $\mu$ for the fsLJ and the WCA model with those
of the hard sphere system, we have also computed the reduced coefficient,
$\mu^{\star}=\mu \sqrt{k_{\rm B}T_{\rm M}/m}$ or, with $m=k_{\rm
B}=1$, $\mu^{\star}=\mu \sqrt{T_{\rm M}}$. It is to be noted that the phase 
behavior of hard spheres is purely entropy-driven and interfacial properties
are solely determined by packing effects. In Figs.~\ref{fig9}b and
\ref{fig10}b, $\mu^\star$ as a function of the melting temperature is
shown for the fsLJ and the WCA model, respectively.  In agreement with a
recent simulation study \cite{pedersen2014}, for both models $\mu^\star$
tends to approach a constant value in the high temperature limit.
The hard-sphere values of $\mu^\star$ , as obtained by Amini and
Laird~\cite{amini2006}, are $\mu_{100}=1.44(7)$, $\mu_{110}=1.10(5)$
and $\mu_{111}=0.64(4)$ for the (100), (110), and (111) orientation,
respectively. The asymptotic values of the fsLJ and the WCA model in
the high pressure/high temperature limit are close to these values,
though not identical.

\begin{figure}
\includegraphics[width=3.0in]{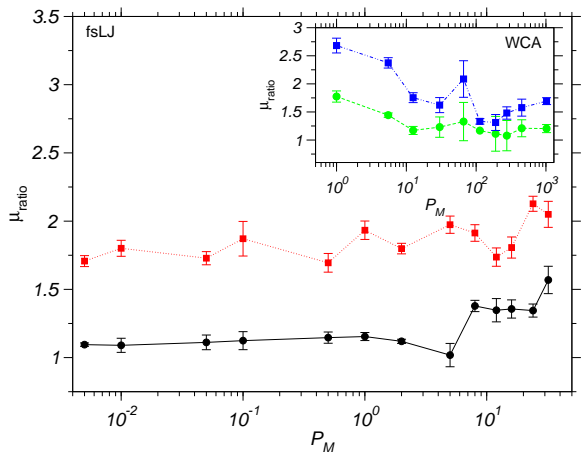}
\caption{(Color online) The ratios $\mu_{\rm 100}/\mu_{\rm 110}$ (circles)
and $\mu_{\rm 100}/\mu_{\rm 111}$ (squares) for the fsLJ system. The
inset shows the same ratios for the WCA system.  \label{fig11}}
\end{figure}

The ratios $\mu_{100}/\mu_{110}$ and $\mu_{100}/\mu_{111}$ reported
in Ref.~\cite{amini2006} for hard spheres are $1.31\pm0.09$ and
$2.25\pm0.18$, respectively.  For a LJ system, a previous simulation study
\cite{pedersen2014} reports the values $1.53$ for $\mu_{100}/\mu_{110}$
and $1.996$ for $\mu_{100}/\mu_{111}$ at $P=0.00254$.
In Fig.~\ref{fig11}, we plot $\mu_{\rm 100}/\mu_{\rm 110}$ and $\mu_{\rm
100}/\mu_{\rm 111}$ as a function of $P_{\rm M}$ for the fsLJ model and
in the inset for the WCA model.  For the fsLJ model, we find that both
ratios are lower than the hard-sphere and LJ values in the pressure
range where the melting temperature hardly changes.  At higher values
of the melting pressure, both ratios tend to similar values as for the
hard-sphere and the LJ system. The ratios $\mu_{\rm 100}/\mu_{\rm 110}$
and $\mu_{\rm 100}/\mu_{\rm 111}$ for the WCA model decrease respectively
from high values of $1.77 \pm 0.099$ and $2.68 \pm 0.13$ at low pressures
and then saturate at values slightly smaller than those for the
hard-sphere and the LJ system.  These results indicate that the specific
nature of the intermolecular potential has a substantial effect on the
magnitude and anisotropy of $\mu$ and cannot be ignored, even though
entropic effects play a dominant role.

Now, we address the question to what extent classical models of
crystal growth can predict the dependence of the interfacial velocity
on undercooling.  The WF model~\cite{wilson1900,frenkel1932}
assumes that crystal growth is an activated process which
is limited by the self-diffusion of the atoms in the liquid
phase. This model leads to the following expression for $v_{\rm
i}$~\cite{jackson1982,jackson2002,hoyt-asta-karma2002}:
\begin{equation}
{v_{\rm i}}^{\rm WF}=
\frac{Ddf}{\varLambda^{2}} {\rm e}^{(-{\Delta S}/{k_B})} 
[1-{\rm e}^{-{\Delta G}/{k_{\rm B}T}}]
\label{eq:wfvi}
\end{equation}
where, $D$ is the self-diffusion constant of the liquid atoms, $d$ the
interplanar spacing between adjacent crystalline layers, $\Lambda$ the
mean free path of a liquid atom, $f$ the probability of a liquid atom to
be attached to a crystal lattice site at the interface, and $\Delta S$ and
$\Delta G$ respectively the differences in entropy and Gibbs free energy per particle
between the crystal and liquid phases. The temperature is assumed to be
below the melting temperature, $T<T_{\rm M}$.  The thermodynamic driving
force for the crystallization is described by the term $e^{(-{\Delta
S}/{k_B})} [1-e^{-{\Delta G}/{k_{\rm B}T}}]$.

The WF model has been shown to predict the crystallization rates of
binary (metallic) systems fairly well~\cite{jackson1982,nascimento2004}.
For one-component systems, however, the WF model tends to underestimate
kinetic growth coefficients.  Therefore, Broughton, Gilmer and Jackson
(BGJ) {\it et al.}~\cite{jackson1982,hoyt-asta-karma2002} predicted an
alternative collision-limited model,
\begin{equation}
{v_{\rm i}}^{\rm BGJ} = 
\frac{df}{\lambda} \sqrt{\frac{3{k_B} T}{m}}  
{\rm e}^{(-{\Delta S}/{k_B})} [1-{\rm e}^{-{\Delta G}/{k_{\rm B}T}}] \, .
\label{eq:bgjvi}
\end{equation}
The BGJ model differs from the WF model in that the prefactor depending on
the self-diffusion constant of the liquid atoms is replaced by a term
containing their thermal velocity such that in the BGJ model the limiting
factor for crystal growth is the thermal velocity with which the atoms
collide rather than the self-diffusion coefficient.  Various works have
compared these two theories to simulation and experimental results and
it has been observed that the collision-limited model is an accurate
predictor of $\mu$ (subject to an appropriate value for the fit parameter
$f$), at least for the (100) and (110) orientation of one-component fcc
systems \cite{hoyt-asta2002,sunastahoyt2004,celestini2002} choosing
the fit parameters \cite{jackson2002} $f=0.27$ and $\lambda=0.15a$
(with $a$ the lattice constant).  In simulations of LJ systems
\cite{jackson1982,jackson2002}, however, the WF model has been shown to
describe the interface velocity corresponding to the (111) orientation
reasonably well.

At small undercoolings, $\Delta G$ is proportional to the undercooling,
$\Delta T$, and the entropy difference can be taken as independent
of temperature. Hence, close to coexistence $1-\exp(-\Delta
G)/{{k_B}T}\approx \frac {L \Delta T}{{k_{B}}TT_{\rm M}}$ and $\Delta S
\approx L/T_{\rm M}$. Inserting these expressions in Eqs.~(\ref{eq:wfvi})
and (\ref{eq:bgjvi}) and using the definition of $\mu$, Eq.~(\ref{eq_vi}),
one obtains the following expressions for $\mu$ corresponding to the
WF model,
\begin{equation}
\mu^{\rm WF}=
\frac{Ddf}{\varLambda^{2}}  
{\rm e}^{(-{L}/{k_B}T_M)} L/(k_{\rm B}TT_{\rm M})
\label{eq:muWF}
\end{equation}
and the BGJ model,
\begin{equation}
\mu^{\rm BGJ}=
\frac{df}{\lambda} \sqrt{\frac{3{k_B} T}{m}} 
{\rm e}^{(-{L}/{k_B}T_M)} L/(k_{\rm B}TT_{\rm M}) \, .
\label{eq:muBGJ}
\end{equation}
Note that since we consider small undercoolings, we can replace the temperature $T$
in Eqs.~(\ref{eq:muWF}) and (\ref{eq:muBGJ}) by the melting
temperature $T_{\rm M}$.

Now, we compare the above predictions for $\mu$ with the simulation
data along the $P-T$ coexistence line.  For both the fsLJ and the
WCA model, $\lambda$ in Eq.~(\ref{eq:muBGJ}) is chosen to be $0.14a$
for comparison with results corresponding to the (100) and (110)
crystal-liquid interfaces.  The parameter $f$ is chosen to be $0.35$
(0.28) and $0.30$ (0.29) for the (100) and (110) orientations of the fsLJ
(WCA) model, respectively.  To compare the WF model, Eq.~(\ref{eq:muWF}),
to the simulation results for the (111) orientation of the fsLJ and the
WCA model,  the parameters $\varLambda$ and $f$ are chosen to be $0.04a$
and $0.27$, respectively.  Note that the same values of these parameters
are used for each point along the coexistence line.

Figures \ref{fig9}a and \ref{fig10}a show reasonable agreement between
theory and simulation.  Using the classical models, the decrease of the
kinetic growth coefficient with increasing melting temperature can be
understood in terms of the decrease of the thermodynamic driving force
along the coexistence line, described by the term $e^{(-{\Delta S}/{k_B})}
[1-e^{-{\Delta G}/{k_{\rm B}T}}] \approx e^{-L/T_{\rm M}}\frac {L \Delta
T}{{k_{B}}TT_{\rm M}}$. The decrease of this term with increasing $T_{\rm
M}$ overpowers the increase of the remaining terms containing $D$ and
$\sqrt{T_{\rm M}}$ leading to slower growth kinetics.

It is to be noted that the values of $f$ in Eq.~(\ref{eq:muBGJ}),
corresponding to the (100) and (110) orientations of the two potentials
that leads to the best agreement with the simulation data, are very
close, indicating that this parameter is independent of the orientation.
In Eq.~(\ref{eq:muBGJ}, $\mu$ corresponding to the different orientations
are proportional to the interplanar spacing $d$ yielding the value
$\mu_{\rm 100}/\mu_{\rm 110}=1.414$.  As shown in Fig.~\ref{fig11},
results from simulations are in reasonable agreement with this value.

%%%
%%%

%%%
\begin{figure}
\includegraphics[width=3.0in]{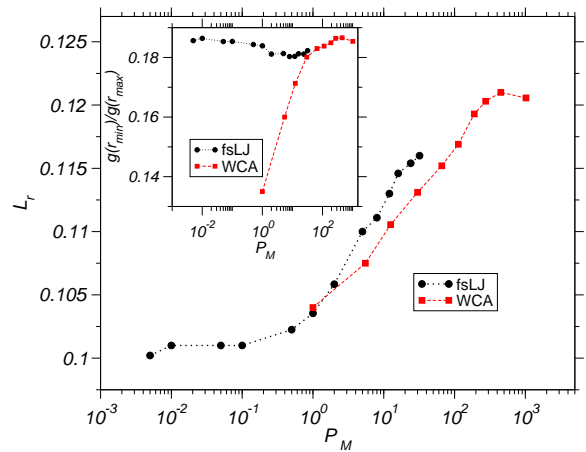}
\caption{(Color online) The Lindemann ratio (in the inset, the RMS
freezing ratio) as a function of pressure for the WCA (red squares)
and fsLJ potentials (black circles). \label{fig12}}
\end{figure}
\subsection{Freezing and melting rules}
\label{subsec:melting_rules}
The FSM method yields simultaneously the $P-T$ coexistence
line and the kinetic growth coefficient, $\mu$. However,
computational approaches such as the capillary fluctuation
method~\cite{monk2002,hoyt-asta-karma2002,hoyt-asta2002,hoyt-asta-karma2003,amini2006,benet2014}
or the equilibrium fluctuation technique~\cite{tepper-briels1997,briels2001,briels2002} require an
accurate determination of the coexistence conditions for obtaining precise
estimates of the kinetic growth coefficient. Computing the coexistence
conditions at several state points to obtain an accurate phase diagram,
is computationally demanding. To reduce the computational effort, it would
be helpful to have a prior idea of the crystal-liquid phase coexistence
region, to avoid carrying out fruitless simulations at regions far away
from the phase boundaries. To this end, there have been a number of
attempts to identify certain empirical rules obeyed by the individual
phases at melting and freezing~\cite{loewen1994,monson2000}. How
accurately such rules are obeyed depends on the nature of the potential
and the coexistence conditions.  Here, we test the validity of three
empirical rules for the two models studied in this work: (i) Lindemann's
melting rule~\cite{lindemann1910}, (ii) the Ravech\'{e}-Mountain-Streett
freezing rule~\cite{rms1974} and (iii) the Hansen-Verlet freezing
criterion~\cite{hv6970}.

According to Lindemann's melting rule, a crystal melts when the root
mean square displacement of crystalline atoms around their ideal lattice
positions is approximately $10\%$ of the nearest-neighbour distance,
$a$. To test this rule, we compute the Lindemann ratio $L_r$, given by
\begin{equation}
L_{\rm r}= \frac{\sqrt{\delta r^2(t)}}{a} \, .
\end{equation}
The values of $L_{\rm r}$, as reported from previous
studies for different model potentials, range roughly from 0.1 to 0.2
\cite{ahmad-sadus2009,saija2006,sadus2009,agrawal-kofke1995}.

To compute $L_{\rm r}$ and also the quantities involved in the other
empirical rules, we have carried out simulations of the bulk crystal
and liquid at the appropriate coexistence temperature and pressure for
systems consisting of $8788$ and $6912$ particles for the WCA and the
fsLJ model, respectively.

Figure~\ref{fig12} shows the variation of $L_{\rm r}$ along the
coexistence line. For both models, it increases from a value of
about $0.10$ at low pressure to one of about $0.12$ at high pressure,
indicating that $L_{\rm r}$ is not invariant along the melting line,
but the closeness of the reported values with Lindemann's prediction
support a qualitative validity of this rule.

%%%
\begin{figure}
\includegraphics[width=3.0in]{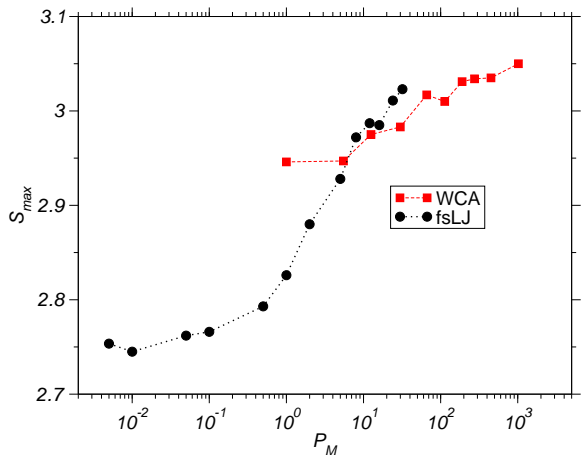}
\caption{(Color online) The maximum of the liquid structure factor,
$S_{\rm max}(k)$, along the coexistence line for the fsLJ (circles)
and WCA (squares) system. \label{fig13}}
\end{figure}

The second empirical rule we test is based on the liquid structure
and known as the Ravech\'{e}-Mountain-Streett (RMS) freezing
criterion. According to this rule, along the freezing line, the radial
distribution function of the liquid obeys the following relation:
\begin{equation}
I=g(r_{\rm min})/g(r_{\rm max})=0.2\pm0.02
\end{equation}
where, $g(r_{\rm min})$ is the first non-zero minimum and $g(r_{\rm max})$
the first maximum of the radial distribution function $g(r)$.  In the
inset of Fig.~\ref{fig12}, we show the RMS freezing criterion along the
coexistence pressure for the WCA and the fsLJ model.  While quantitatively
our results differ from the RMS rule, the data corresponding to the fsLJ
model, which hover around $0.185$ at low pressures (between $0.005$ and
$1.0$) and around $0.180$ at larger values of $P_{\rm M}$, point to a
rough invariance of this quantity along the coexistence line. However,
at low pressures ($P_{\rm M}\leq 30$), $I$ corresponding to the
WCA potential is much smaller than suggested by the RMS criterion, while
at larger pressures, it saturates around value of about $0.186$, close
to the value obtained for the fsLJ model. A similar variation of the
RMS freezing criterion along the coexistence line was also observed in
an earlier work~\cite{sadus2009}.

The Hansen-Verlet freezing rule states that the first peak of the liquid
structure factor attains the value $S_{\rm max}(k)=2.85$ and remains
invariant along the freezing curve. Hansen and Verlet postulated this
rule based on observations corresponding to the LJ potential. Later,
Agrawal and Kofke~\cite{agrawal-kofke1995} reported a $10\%$ increase
of this quantity along the coexistence line for the LJ potential.
We computed the structure factor directly from the coordinates of
the particles and not via a Fourier transform of the pair correlation
function \cite{binder2004}. Our results (Fig.~\ref{fig13}) for the two systems indicate a
systematic increase of $S_{\rm max}(k)$ along the freezing curve. While
$S_{\rm max}$ corresponding to the fsLJ system increases from a minimum
of $2.74$ to $3.02$ at the highest pressure, that for the WCA potential
remains above the value predicted by the Hansen-Verlet freezing criterion
at all the considered coexistence pressures. This indicates that the
Hansen-Verlet criteria can at most be used as a rule of thumb principle
to indicate how close the system is to the actual freezing temperature.

\section{Conclusion}
\label{subsec:conc}
We have computed the crystal-melt kinetic growth coefficient from
molecular dynamics simulation via the free solidification method. Our
results are consistent with previous works as regards the magnitude and
anisotropy of $\mu$ is concerned.  The variation of the kinetic growth
coefficient along coexistence indicates that the slowing down of the
crystal growth is primarily a temperature effect.  Changing the pressure
by two orders of magnitude in case of the fsLJ potential does not change
the magnitude of the kinetic growth coefficient in the coexistence region
where the melting temperature remains almost unchanged.

Similarity of values between Lennard-Jones and hard-sphere systems indicate
that packing effects play a dominant role in describing the growth
kinetics. However, the specific nature of the interaction potential cannot
be ignored.  Classical theories of crystallization models predict values
of $\mu$ in reasonably good agreement with simulation results, subject to
an appropriate value of a fit parameter.  In the high-temperature and
high-pressure limit, the crystal growth kinetics of the fsLJ and WCA
potentials are similar to each other. In this limit, the coexistence
properties of the two systems approach that of the purely repulsive,
inverse twelfth-power potential.  Various melting and freezing rules have
been investigated and while they roughly indicate the coexistence region,
they are not quantitatively accurate.

{\bf Acknowledgments:} The authors thank Roberto Rozas for useful discussions.
RB thanks DAAD for financial support and DLR, Cologne, for computational
and research facilities during the initial phase of this project.
The authors acknowledge financial support from the German DFG in the
framework of the M-era.Net project ``ANPHASES''.

\end{document}